\documentclass[12pt]{article}
\usepackage{epsf,amsmath,amscd,amssymb,graphicx}
\usepackage{amsfonts}
\begin{document}

\thispagestyle{empty}

\begin{flushright}
TIT/HEP--514 \\
{\tt hep-th/0311210} \\
November, 2003
\end{flushright}
\vspace{3mm}
\begin{center}
{\Large
{\bf An exact BPS wall solution} 
\\
\vspace{2mm}
{\bf in Five Dimensional Supergravity}

\vspace{7mm}

\normalsize

  {\large \bf 
  Masato~Arai~$^{a}$}
\footnote{Talk given at the International Workshop 
 ``Supersymmetries and Quantum 
 Symmetries'' (SQS 03), BLTP, JINR, Dubna, Russia, July 24-29, 2003.}
\footnote{\it  e-mail address: 
arai@fzu.cz
},
 {\large \bf 
Shigeo~Fujita~$^{b}$}
\footnote{\it  e-mail address: 
fujita@th.phys.titech.ac.jp
},
  {\large \bf 
  Masashi~Naganuma~$^{b}$}
\footnote{\it  e-mail address: 
naganuma@th.phys.titech.ac.jp
},  

~and~~  {\large \bf 
Norisuke~Sakai~$^{b}$}
\footnote{\it  e-mail address: 
nsakai@th.phys.titech.ac.jp
} 

\vskip 1.5em

{ \it $^{a}$
  Institute of Physics, AS CR, 
  182 21, Praha 8, Czech Republic \\
  and \\
  $^{b}$Department of Physics, Tokyo Institute of 
  Technology \\
  Tokyo 152-8551, JAPAN 
}
\\
\vspace{5mm}
%
{\bf Abstract}\\[5mm]
{\parbox{13cm}{\hspace{5mm}
In five-dimensional supergravity, 
an exact solution of BPS wall is found 
for a gravitational deformation of 
the massive Eguchi-Hanson nonlinear sigma model. 
The warp factor decreases for both infinities of the extra dimension. 
Thin wall limit gives the Randall-Sundrum model without fine-tuning 
of input parameters. 
We also obtain wall solutions with warp factors which are flat or 
increasing in one side, by varying 
a deformation parameter of 
the potential. 
}}}
\end{center}
\section{Introduction} 
One of the most interesting models in the brane-world scenario is 
 given by Randall and Sundrum, where 
 the localization of four-dimensional 
 graviton \cite{RS2} has been obtained 
 by a spacetime metric 
 containing a warp factor $e^{2U(y)}$ which decreases 
 exponentially for both infinities of the extra dimension 
$y \rightarrow \pm \infty$ 
\begin{eqnarray}
  ds^2=g_{\mu \nu}dx^\mu dx^\nu = e^{2U(y)}\eta_{mn}dx^mdx^n + dy^2,
\label{5dmetric}
\end{eqnarray}
where $\mu, \nu = 0,..,4$, $m, n = 0,1,3,4$ and $y\equiv x^2$. 
They had to introduce both a bulk cosmological constant 
 and a boundary cosmological constant, which have 
 to be fine-tuned each other. 

This scenario is based on the assumption of the 
 existence of delta functional domain wall.
Thus, it would be nice to obtain the domain wall as a classical
 solution in some field theory in a phenomenological point of view.
Studies of domain wall solutions in gauged supergravity theories 
 in five dimensions revealed that hypermultiplets are needed 
\cite{KalloshLinde} to obtain 
 warp factors decreasing for both infinities $y \rightarrow \pm \infty$ 
 (infra-red (IR) fixed points in AdS/CFT correspondence \cite{AdS/CFT}). 
It has been shown that the target space of hypermultiplets in 
 five-dimensional supergravity theory must be quaternionic K\"ahler (QK)
 manifolds \cite{BaggerWitten}.
Further, in order to obtain domain wall of hypermultiplets, 
 mass terms of them are needed.
Domain walls in massive QK nonlinear sigma models (NLSM) 
 in supergravity theories have been studied using 
 mostly homogeneous target manifolds. 
Unfortunately, supersymmetric (SUSY) vacua in 
 homogeneous target manifolds are not truly IR critical points, 
 but can only be saddle points with some IR directions 
 \cite{Alekseevsky}, \cite{BC2}. 
Inhomogeneous manifolds and a wall solution have also been 
 constructed \cite{Beh-Dall}, \cite{Lazaroiu}. 
However, 
 these manifolds do not allow a limit of weak 
 gravitational coupling.

The purpose of this paper is to present an exact BPS domain wall 
 solution in five-dimensional supergravity coupled with hypermultiplets 
 (and vector multiplets).
Our strategy to construct the model is to deform the NLSM 
 in SUSY theory having domain wall solution 
 to the model with gravity.
Massive hyper-K\"ahler NSLMs without gravity 
 in four dimensions have been constructed 
 in harmonic superspace as well as in ${\cal N}=1$ superfield 
 formulation \cite{ANNS}, and have yielded the domain wall 
 solution for the Eguchi-Hanson (EH) manifold \cite{Eguchi}.
Inspired by this solution,
 we deform this model into five-dimensional supergravity model
 and we consider the BPS domain wall solution.
We also discuss a limit of weak gravitational
 coupling.
This paper is based on our paper \cite{ANFS} where complete analysis
 and references are found.

\section{Bosonic action of our model in 5D Supergravity}
To find a gravitational deformation of the NLSM
 with EH target manifold, 
 we use the recently obtained off-shell formulation 
 of five-dimensional 
 supergravity (tensor calculus) \cite{Fujita-Ohashi}, \cite{FKO} 
 combined with the quotient method via a vector multiplet without 
 kinetic term and the massive deformation (central charge extension).
\footnote{ 
 We adopt the conventions of Ref. \cite{Fujita-Ohashi} except 
 the sign of our metric $\eta_{\mu\nu}= diag(-1,+1,+1,+1,+1)$. 
 This induces a change of
 Dirac matrices 
 and the form of SUSY transformation of fermion.
 }
We start with the system of 
 a Weyl multiplet, three hypermultiplets and two $U(1)$ vector multiplets. 
One of the two 
 vector multiplets has no kinetic term and plays the role of a Lagrange 
 multiplier for hypermultiplets to obtain a curved target manifold. 
The other vector multiplet serves 
 to give mass terms for hypermultiplets. 

After integrating out a part of the auxiliary fields 
 by their on-shell conditions in 
 the off-shell supergravity action \cite{FKO}, 
 we obtain the bosonic part of the action 
 for our model 
\begin{eqnarray}
   e^{-1}{\cal L}
   &\!\!\!=&\!\!\!
  -\frac{1}{2\kappa^2}R-\frac{1}{4}
  \left(\partial_\mu W_\nu^0-\partial_\nu W_\mu^0 \right)
  \left(\partial^\mu W^{0\nu}-\partial^{\nu} W^{0\mu} \right)
\nonumber \\
   &\!\!\!{}&\!\!\! 
-\nabla^a{\cal A}_i^{\beta}d_\beta{}^\alpha\nabla_a{\cal A}^i_\alpha
    - \kappa^2[{\cal A}^\beta{}_id_\beta{}^\alpha\nabla_a{\cal A}^j_\alpha]^2 
       \nonumber \\
   &\!\!\!{}&\!\!\! -\left[-{\cal A}_i{}^\gamma d_\gamma{}^\alpha 
           (g_0M^0t_0+ M^1t_1)^2{}_\alpha{}^\beta {\cal A}^i_\beta
          -\frac{\kappa^2}{12}(g_0M^0)^2
           (2{\cal A}_\alpha^{(i}d^\alpha{}_\gamma
                (t_0)^{\gamma \beta}{\cal A}^{j)}_\beta)^2 \right], 
\label{SUGRA1} \\
\nabla_\mu{\cal A}_i^{\alpha}
  &=& \partial_\mu {\cal A}_i^\alpha
      -(g_0W^0_\mu t_0+ W^1_\mu t_1)^\alpha{}_\beta {\cal A}_i^\beta ,~~~~
{\cal A}^i{}_\alpha \equiv \epsilon^{ij}{\cal A}_j{}^\beta 
\rho_{\beta \alpha} = - ({\cal A}_i{}^\alpha)^* , \nonumber
\end{eqnarray}
where $d_\alpha{}^\beta = diag(1,1,-1,-1,-1,-1)$, 
 $\kappa$ is the five-dimensional gravitational coupling, 
 ${\cal A}_i^\alpha, \; i=1,2, \; \alpha=1,\dots,6$ 
 are the scalars in hypermultiplets, and 
 $W_\mu^0$ ($W_\mu^1$), 
 $M^0$ ($M^1$) and $t_0$ ($t_1$) are vector fields, scalar fields 
 and generators of the $U(1)$ vector multiplets 
 with (without) a kinetic term. 
The gauge coupling of $W_\mu^0$ is denoted by $g_0$. 
Another gauge coupling $g_1$ is absorbed into a normalization of 
 $W_\mu^1$ in order to drop the kinetic term by taking 
 $g_1 \rightarrow \infty$. 
Hypermultiplet scalars are subject to two kinds of constraints
\begin{eqnarray}
  {\cal A}^2={\cal A}^\beta_id_\beta{}^\alpha{\cal A}^i_\alpha 
   = -2 \kappa^{-2},~~~~~ 
  {1 \over g_1^2}{\cal Y}_1^{ij}
  &\equiv & 2{\cal A}_\alpha^{(i}d^\alpha{}_\gamma
                (t_1)^{\gamma \beta}{\cal A}^{j)}_\beta = 0 .
\label{const}
\end{eqnarray}
The first constraint comes from the gauge fixing 
 of dilatation, and 
 make target space of hypermultiplets to be a non-compact version of 
 quaternionic projective space, $\frac{Sp(2,1)}{Sp(2)\times Sp(1)}$, 
 combined with the gauge fixing of 
 $SU(2)_R$ symmetry. 
The second constraint is required by the on-shell condition 
 of auxiliary fields of the $U(1)$ vector multiplet without kinetic term, 
 and corresponds to the constraint for the EH target space 
 in the limit of $\kappa\to 0$. 

The third line of (\ref{SUGRA1}) is a scalar potential. 
The scalar $M^0$ is fixed as 
 $(M^0)^2=\frac{3}{2}\kappa^{-2}$ 
 from the requirement of canonical normalizations of the Einstein-Hilbert 
 term and the kinetic term of 
 the gravi-photon $W^0_\mu$ for Poincar\'e supergravity. 
The scalar $M^1$ without kinetic term is a Lagrange multiplier, 
 and is found to be 
\begin{eqnarray}
 M^1=-\frac{{\cal A}_i^\gamma d_\gamma{}^\alpha 
     (t_0t_1)_\alpha{}^\beta{\cal A}^i{}_\beta}
     {{\cal A}_i{}^\gamma d_\gamma{}^\alpha(t_1)^2_\alpha{}^\beta
     {\cal A}^i{}_\beta}g_0M^0 .
\end{eqnarray}

Here we introduce
 two two-component complex fields $\phi_1$ and $\phi_2$ 
 to parametrize ${\cal A}_i{}^\alpha$ by a matrix with $i=1, 2$ as rows 
 and $\alpha=1,\dots,6$ as columns 
\begin{eqnarray}
 {\cal A}_i{}^\alpha \equiv \frac{1}{\kappa}\bar{\cal A}^{-1/2}\left(
\begin{array}{cccc}
1 & 0 & \kappa\phi_1 & -\kappa\phi_2^* \\
0 & 1 & \kappa\phi_2 & \kappa\phi_1^* 
\end{array} \right) \label{CFbasis} 
\end{eqnarray}
satisfying the first constraint in (\ref{const}) by 
taking $\bar{\cal A}=1-\kappa^2(|\phi_1|^2+|\phi_2|^2)$. 
In this basis, 
 we can choose two $U(1)$ generators as 
${t_1}^{\alpha}_{~\beta}=diag(i\alpha,-i\alpha,i,i,-i,-i)$ and 
${t_0}^{\alpha}_{~\beta}=diag(ia\alpha,-ia\alpha,-i,i,i,-i)$,
where $\alpha$ and $a$ are real parameters.
The parameter $\alpha $ in $t_1$ makes target manifold 
 inhomogeneous generally through the second constraint 
 in (\ref{const}), 
 and a special case of $\alpha=1$ corresponds to a homogeneous 
 manifold of $SU(2,1)/U(2)$ \cite{BS}.
Here we define $\alpha \equiv \kappa^2\Lambda^3$,
 where $\Lambda$ is a real parameter of unit mass dimension. 

\begin{figure}
\begin{center}
\leavevmode
\begin{eqnarray*}
\begin{array}{cc}
  \epsfxsize=6cm
  \epsfysize=4.5cm
\epsfbox{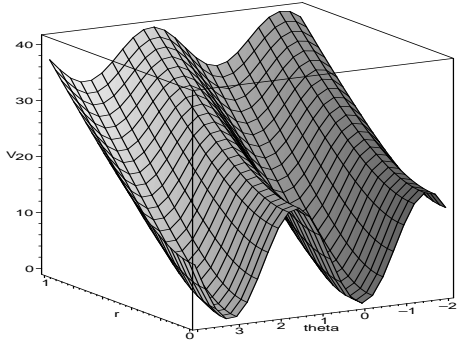} & 
  \epsfxsize=6cm
  \epsfysize=4.5cm
\hspace*{1.5cm}
\epsfbox{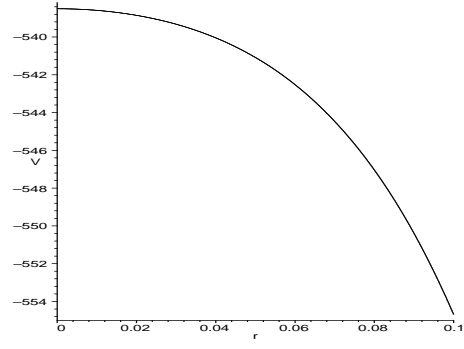} 
\\
\mbox{\footnotesize (a) Potential for $a=0$ at $\kappa = 0.1$} & 
\hspace*{1.5cm}
\mbox{\footnotesize (b) Potential at $\kappa^2\Lambda^3 = 0.9$}
\end{array} 
\end{eqnarray*} 
\caption{Discrete vacua. 
Parameters are taken to be $(g_0M^0, \Lambda) = (3, 1)$.}
\label{pot-plot}
\end{center}
\end{figure}

In order to solve constraints
 (\ref{const}) in terms of (\ref{CFbasis}),
 we introduce the spherical coordinates as
\begin{eqnarray}
\begin{array}{ll}
  \phi_1^1=g(r)\cos(\frac{\theta}{2})\exp(\frac{i}{2}(\Psi+\Phi)), &
\phi_1^2=g(r)\sin(\frac{\theta}{2})\exp(\frac{i}{2}(\Psi-\Phi)), 
\vspace{1mm} \\
\phi_2^1=f(r)\sin(\frac{\theta}{2})\exp(-\frac{i}{2}(\Psi-\Phi)), &
\phi_2^2=-f(r)\cos(\frac{\theta}{2})\exp(-\frac{i}{2}(\Psi+\Phi)).
\end{array}
\label{Constsolve}
\end{eqnarray}
Here we set 
\begin{eqnarray}
  f(r)^2=\frac{1}{2}(-\Lambda^3+\sqrt{4r^2+\Lambda^6}), \;\;
  g(r)^2=\frac{1}{2}(\Lambda^3+\sqrt{4r^2+\Lambda^6})
\end{eqnarray}
in order to satisfy (\ref{const}).

Substituting the solution (\ref{Constsolve}) into the action,
 it can be described by independent variables. 
The target metric of the kinetic term 
 is found to be a quaternionic extension of the
 EH metric \cite{Galicki},~\cite{IvanovValent}.
Since the metric is Einstein, the Weyl tensor is anti-selfdual and 
 the scalar curvature is negative $R=-24\kappa^2$, 
 it is locally a quaternionic manifold \cite{BaggerWitten} 
 for any values of $\kappa \ne 0$.

The scalar potential part in (\ref{SUGRA1}) becomes a function of
 fields $r,~\theta$ and depends on the parameter $a$ and the gravitational 
 coupling $\kappa$, i.e. $V=V(r,\theta,a,\kappa)$.
Here we do not show the explicit form but only show the plot of 
 the potential.
Fig.~1-(a) shows 3D plot of the potential and it is found
 there exist two vacua 
 at $(r,\theta)=(0,0),(0,\pi)$ as local minima. 
These two vacua become saddle points with an unstable direction 
 along $r$ for $\kappa^2\Lambda^3 > 3/4$ 
 for $a=0$. 
Fig.~1-(b) shows a typical unstable behavior of potential 
 at $\kappa^2\Lambda^3 =0.9$, which is close to 
 $\kappa^2\Lambda^3 =1$, where the target space of hypermultiplets
 becomes a homogeneous space of $SU(2,1)/U(2)$. 
For $a\ne 0$, potential takes different values at these two vacua. 

\section{BPS equation and the solution} 
Instead of solving Einstein equations directly, we solve BPS equations  
 to obtain a classical solution conserving a half of SUSY. 
Since we consider bosonic configurations, we need to examine 
 the on-shell SUSY transformation 
 of gravitino and hyperino \cite{Fujita-Ohashi}
\begin{eqnarray}
 \delta_\varepsilon \psi^i_\mu 
  &=& {\cal D}_\mu \varepsilon ^i 
        - \frac{\kappa^2}{6}M_0{\cal Y}_0^{i}{}_j
          \gamma_\mu \varepsilon ^j ,
\label{gravitino1} \\
  \delta_\varepsilon \zeta^\alpha 
   &=& -{\cal D}_\mu {\cal A}^\alpha{}_j\gamma^\mu \varepsilon ^j 
       -( M^1t_1+g_0M^0t_0)^\alpha{}_\beta {\cal A}^\beta{}_j \epsilon^j 
       + \frac{\kappa^2}{2}{\cal A}_j{}^\alpha
        M_0{\cal Y}_0^j{}_k \varepsilon ^k ,
  \label{hyperino1}
\end{eqnarray} 
where 
\begin{eqnarray}
  {\cal D}_\mu \varepsilon^i 
   &=& \left(\partial_\mu -\frac{1}{4}\gamma_{ab}\omega_\mu^{ab}\right)
       \varepsilon^i
        - \kappa^2V_\mu{}^i{}_j\varepsilon^j ,  \\
  {\cal D}_\mu {\cal A}_i^\alpha 
   &=& \partial_\mu {\cal A}_i^\alpha +\kappa^2V_\mu{}_i{}^j{\cal A}_j^\alpha 
         -  W^1_\mu t_1^\alpha{}_\beta {\cal A}_i^\beta ,
\label{cov-deriv} \\
  {\cal Y}_0^{ij} &=& 2{\cal A}_\alpha^{(i}d^\alpha{}_\gamma
                (g_0t_0)^{\gamma \beta}{\cal A}^{j)}_\beta, \quad 
  V_\mu^{ij} = 
   - {\cal A}^{\gamma (i}d_\gamma^{~\alpha}\nabla_\mu{\cal A}^{j)}_\alpha .
\end{eqnarray}

Let us require vanishing of the SUSY variation of gravitino and hyperino 
 to preserve four SUSY specified by 
\begin{eqnarray}
  \gamma^y \varepsilon^i(y) = i \tau_3^{i}{}_j \varepsilon^j(y),
\label{1/2SUSY}
\end{eqnarray}
where $\tau_3$ is one of the Pauli matrix.
Substituting this condition (\ref{1/2SUSY})
 and the metric ansatz (\ref{5dmetric}) 
 into (\ref{gravitino1}) and (\ref{hyperino1}), we obtain BPS equation.
We can solve it 
 in the spherical coordinate (\ref{Constsolve}). 
The wall solution interpolating between the two vacua
 $(r, \theta)=(0,0),(0,\pi)$ is obtained from (\ref{hyperino1})
\begin{eqnarray}
 r=0, \;\; \cos \theta = \tanh \left(2g_0M^0(y-y_0)\right),\;\; 
\Phi=\varphi_0,
\label{BPSsol-spherical}
\end{eqnarray}
with $\Psi$ undetermined, and $y_0$ and $\varphi_0$ are constants.
Here we take the boundary condition $r=0$ at $y=-\infty$.
Using (\ref{BPSsol-spherical}),
 we obtain the BPS solution of the warp factor and the Killing 
 spinor from (\ref{gravitino1}) 
\begin{eqnarray}
 &&U(y) = -\frac{\kappa^2\Lambda^3}{3(1-\kappa^2\Lambda^3)}
        \left[ \ln \{\cosh \left(2g_0M^0(y-y_0)\right)\} 
           + 2ag_0M^0(y-y_0)\right], \label{Warp-sol} \\
 &&\varepsilon^i(y)\equiv e^{U(y)/2}\tilde{\varepsilon}^i, \quad 
 \gamma^y \tilde{\varepsilon}^i = i\tau_{3}^{i}{}_j\tilde{\varepsilon}^j,
\label{killing-sol}
\end{eqnarray}
where $\tilde{\varepsilon}^i$ is a constant spinor.

The warp factor $e^{2U(y)}$ of this solution 
 decreases exponentially for both infinities $y\rightarrow \pm \infty$ 
 for $|a|<1$ (see Fig.~2) similarly to the case of the bulk AdS space. 
The case of
 $|a|=1$ becomes the wall solutions interpolating 
 between AdS and flat Minkowski vacua. 
On the other hand, warp factor increases exponentially either one of the 
 infinities for $|a|>1$. 
 {}Following the AdS/CFT conjecture, a vacuum reached by a decreasing 
 (increasing) warp factor corresponds to IR (UV) fixed point of a 
 four-dimensional field theory \cite{AdS/CFT}. 
Our BPS wall solutions interpolate two IR fixed points for 
 $|a|<1$. 
The wall solutions for $|a|>1$ interpolate one IR and one UV fixed points 
 which cannot realize the warped 
 extra dimension, but should be related to a Renormalization 
 Group (RG) flow.
The family of our BPS solutions contains a parameter $a$ interpolating 
 between three classes of field theories : 
 one with two IR fixed points ($|a|<1$), 
 another with one IR and one UV fixed point ($|a|>1$), 
 and one with one IR fixed point and flat space ($|a|=1$). 
We find it remarkable that a 
 single family of models can realize all these possibilities 
 as we change a parameter. 

\begin{figure}[t]
\begin{center}
\leavevmode
\begin{eqnarray*}
\begin{array}{ccc}
  \epsfxsize=4.5cm
  \epsfysize=3.5cm
\epsfbox{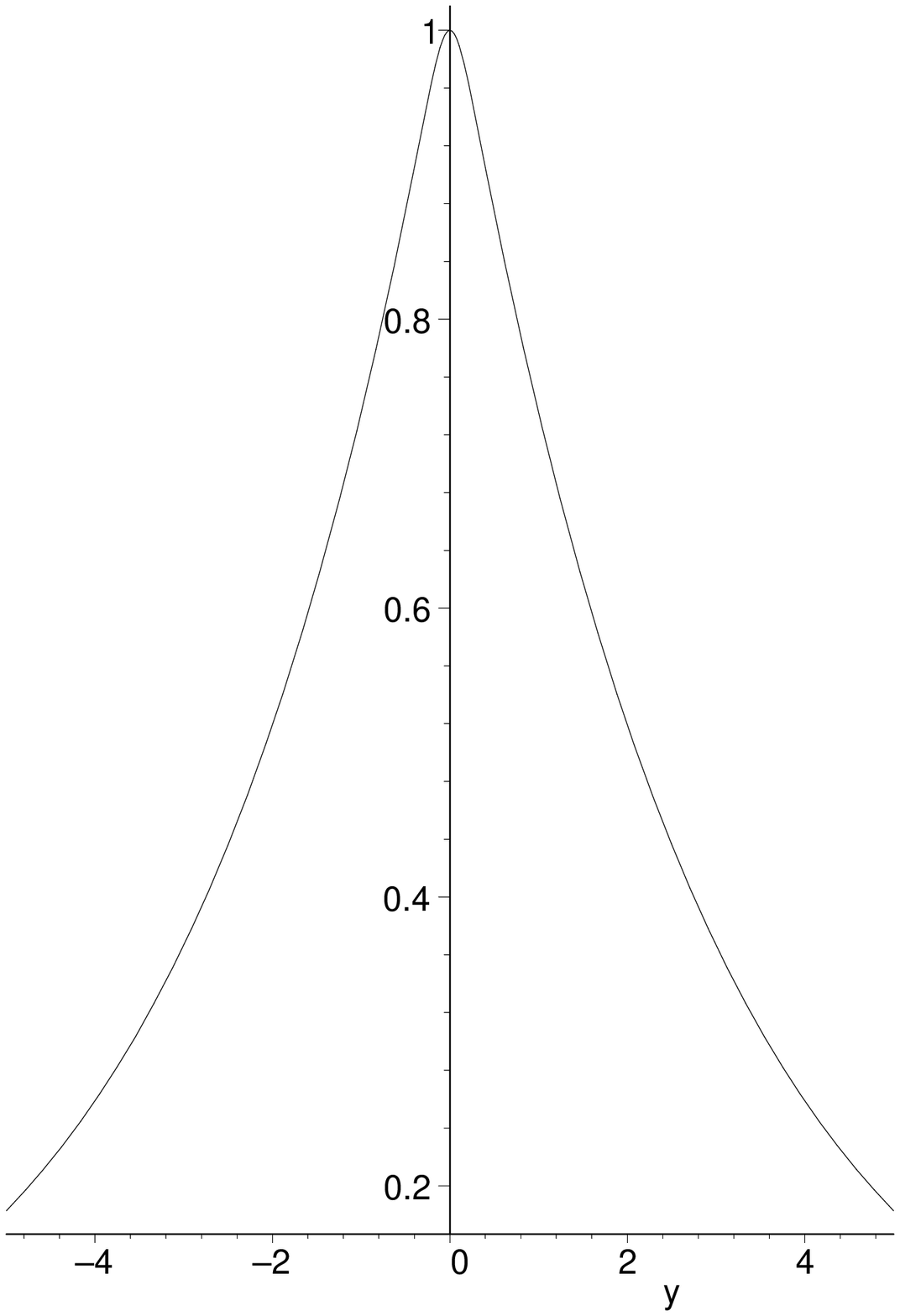} & 
  \epsfxsize=4.5cm
  \epsfysize=3.5cm
\hspace*{1cm}
\epsfbox{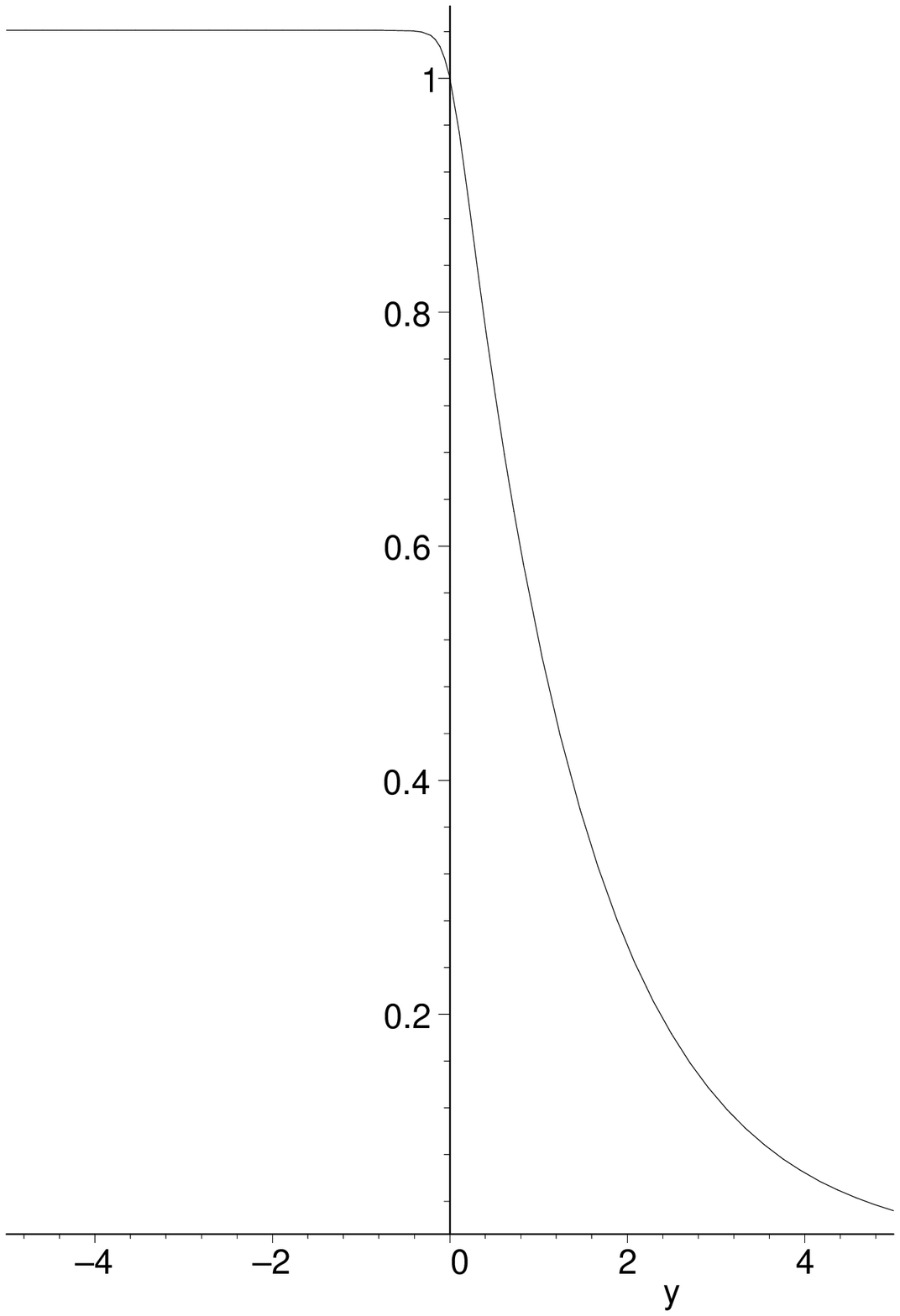} %
  \epsfxsize=4.5cm
  \epsfysize=3.5cm
\hspace*{1cm}
\epsfbox{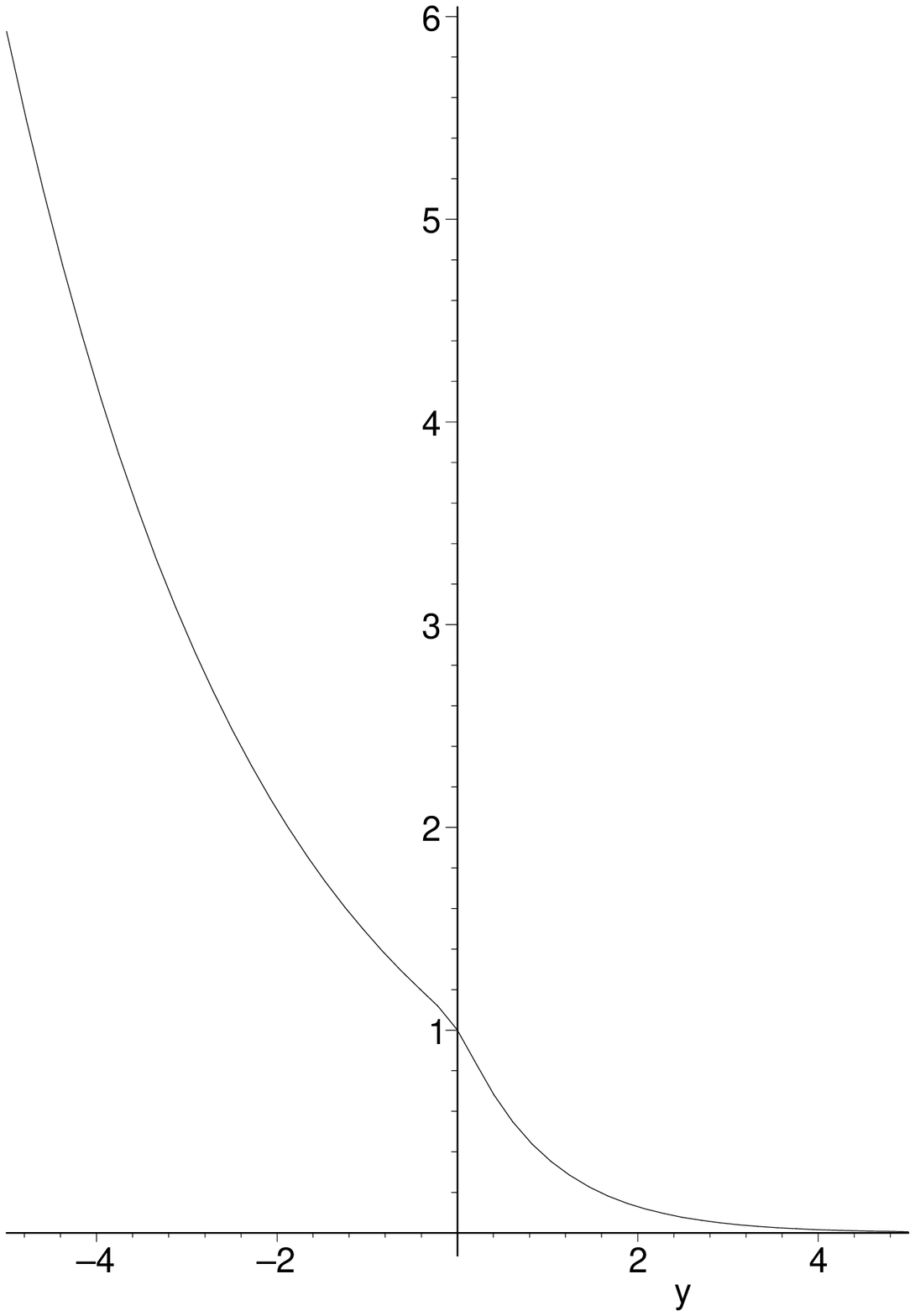} 
\hspace{3cm}
\\
\hspace{0cm}\mbox{\footnotesize (a) a=0} &
\hspace{-8cm}\mbox{\footnotesize (b) a=1}  &
\hspace{-6cm}\mbox{\footnotesize (c) a=2} 
\end{array} 
\end{eqnarray*} 
\caption{Profile of warped metric. 
Parameters are taken to be $(g_0M^0, \Lambda, \kappa) = (3, 2, 0.1)$}
\end{center}
\end{figure}

We can obtain a thin wall limit by taking $g_0M^0 \rightarrow \infty$ 
 and $\Lambda \rightarrow 0$ with $g_0M^0\Lambda^3$, $\kappa$, 
 and $a$ fixed. 
Substituting the solutions (\ref{BPSsol-spherical}) and (\ref{Warp-sol}) 
 to the Lagrangian of hypermultiplets and taking the thin wall limit, 
 we obtain for $y_0=0$ 
\begin{eqnarray}
-\frac{1}{2\kappa^2}R + e^{-1}{\cal L}_{kin} 
    + e^{-1}{\cal L}_{pot}
 \to -\frac{1}{2\kappa^2}R 
  - \Lambda_c^\pm(y) - T_w \delta (y) ,
\end{eqnarray}
where $T_w$ is a cosmological constant (wall tension) 
 $\Lambda_c^{+}, (\Lambda_c^{-})$ is bulk cosmological constant 
 for $y<0 (y>0)$ as 
\begin{eqnarray}
  T_w = 4(g_0M^0\Lambda^3),
 \qquad 
   \Lambda_c^{\pm} 
      = -\frac{8\kappa^2(g_0M^0\Lambda^3)^2}{3}(1 \pm a)^2 .
\end{eqnarray}
We find that  our BPS solution for $a=0$ 
automatically satisfies 
 the fine-tuning condition $\sqrt{-\Lambda_c}=\frac{\kappa}{\sqrt{6}}T_w$ 
 of the Randall-Sundrum model between 
$T_w$ and 
$\Lambda_c$, 
as a result of combined dynamics of scalar field and 
gravity. 
In terms of the asymptotic linear exponent $c$ of the warp factor 
$ U\sim - c |y-y_0|, \; c \equiv 2\kappa^2(g_0M^0\Lambda^3)/3$ for 
$|y-y_0|\to \infty$,  
the wall tension 
$T_w =24c/(4\kappa^2)$,  and cosmological constant 
 $\Lambda_c = -24 c^2/(4\kappa^2)$ 
satisfy precisely the same 
relation as in Ref.~\cite{RS2} 
(with $M_p^3\equiv (4\kappa^2)^{-1}$). 
Therefore we have realized the single-wall Randall-Sundrum model 
as a thin-wall limit of our solution of the coupled scalar-gravity theory, 
instead of an artificial boundary cosmological constant put 
at an orbifold point. 

Finally, we discuss the properties of our model and solution 
 in the weak gravity limit, which is defined by taking 
 the limit of $\kappa \to 0$ with $g_0M^0\equiv \bar{M}$ 
 held fixed. 
In the limit, we find that the action (\ref{SUGRA1})
 in terms of (\ref{CFbasis}) 
 reduces to the five-dimensional version of the massive
 NLSM with the EH target metric in the basis
 in Ref. \cite{CF}. 
The wall solution for $\kappa=0$ is the five-dimensional version of 
 the kink solution in Ref. \cite{ANNS}. 
Their solution is exactly identical to our solution 
 (\ref{BPSsol-spherical}) obtained for finite $\kappa$. 
It is very interesting that
 BPS solution for the hypermultiplet
 in the global SUSY model coincides with 
 that in the corresponding supergravity. 
This mysterious coincidence has also appeared in the analytic 
 solution in a four-dimensional ${\cal N}=1$ supergravity 
 model \cite{EMSS}.  



\begin{thebibliography}{100}
 \bibitem{RS2} L.~Randall and R.~Sundrum, 
              {\it Phys.\ Rev.\ Lett.}~{\bf 83} (1999) 4690,
              [hep-th/9906064].
 \bibitem{KalloshLinde} R.~Kallosh and A.~Linde, 
             {\em JHEP} {\bf 0002} (2000) 005, [hep-th/0001071];
              K.~Behrndt, C.~Herrmann, J.~Louis and S.~Thomas, 
             {\em JHEP} {\bf 0101} (2001) 011, [hep-th/0008112].
 \bibitem{AdS/CFT} J.~Maldacena,
              {\it Adv.\ Theor.\ Math.\ Phys. }{\bf 2}~(1998)~231,
              [hep-th/9711200];
               O.~Aharony, S.S.~Gubser, J.~Maldacena, H.~Ooguri and Y.~Oz, 
              {\it Phys.\ Rep.}~{\bf 323} (2000) 183,
              [hep-th/9912001].
 \bibitem{BaggerWitten} J.~Bagger and E.~Witten, 
                {\it Nucl.\ Phys.}~{\bf B222} (1983) 1. 
 \bibitem{Alekseevsky} D.V.~Alekseevsky, V.~Cort\'es, C.~Devchand 
                and A.~Van~Proeyen, 
               {\it Comm. Math. Phys. }{\bf 238} (2003) 525,  
               [hep-th/0109094]. 
 \bibitem{BC2} K.~Behrndt and M.~Cvetic,
             {\it Phys.\ Rev.}~{\bf D65} (2002) 126007,
             [hep-th/0201272].
 \bibitem{Beh-Dall} K.~Behrndt and G.~Dall'Agata, 
             {\it Nucl.\ Phys.} {\bf B627} (2002) 357, 
             [hep-th/0112136].
 \bibitem{Lazaroiu}  L.~Anguelova and C.I.~Lazaroiu,
             {\it JHEP} {\bf 0209} (2002) 053, 
             [hep-th/0208154].
 \bibitem{ANNS} M.~Arai, M.~Naganuma, M.~Nitta and N.~Sakai, 
                {\it Nucl. Phys.}~{\bf B652} (2003) 35,
                [hep-th/0211103].
 \bibitem{Eguchi} T.~Eguchi and A.J.~Hanson,
                {\it Phys.\ Lett.~}{\bf 74B}~(1978)~249;
                {\it Ann.\ Phys.~}{\bf 120}~(1979)~82.
 \bibitem{ANFS}M.~Arai, S.~Fujita, M.~Naganuma and N.~Sakai,
               {\it Phys.\ Lett.\ }{\bf B556} (2003) 192,
               [hep-th/0212175].
 \bibitem{Fujita-Ohashi} T.~Kugo and K.~Ohashi,
                {\it Prog.\ Theor.\ Phys.}~{\bf 105} (2001) 323,
                [hep-ph/0010288];
                 T.~Fujita and K.~Ohashi, 
                {\it Prog.\ Theor.\ Phys.}~{\bf 106} (2001) 221,
                [hep-th/0104130].
 \bibitem{FKO} T.~Fujita, T.~Kugo and K.~Ohashi,
                {\it Prog.\ Theor.\ Phys.}~{\bf 106} (2001) 671,
                [hep-th/0106051]. 
 \bibitem{BS} P.~Breitenlohner and M.F.~Sohnius,
                {\it Nucl.\ Phys.}~{\bf B187} (1981) 409. 
 \bibitem{Galicki} K.~Galicki,
                {\it Nucl.\ Phys.}~{\bf B271} (1986) 402. 
 \bibitem{IvanovValent} E.~Ivanov and G.~Valent, 
                {\it Nucl.\ Phys.}~{\bf B576} (2000) 543, 
                [hep-th/0001165]. 
 \bibitem{CF} T.L.~Curtright and D.Z.~Freedman,
                {\it Phys.\ Lett.}~{\bf 90B} (1980) 71. 
 \bibitem{EMSS} M.~Eto, N.~Maru, N.~Sakai and T.~Sakata,
             {\it Phys.\ Lett.}~{\bf B553} (2003) 87, 
             [hep-th/0208127].
\end{thebibliography}
\end{document}